\documentclass[aps,prb,amsmath,
groupedaddress]{revtex4}

\usepackage{graphicx}
\begin{document}

\title{Spin Transport in Two Dimensional Hopping Systems}
\author{T. Damker}
\email{thomas.damker@physik.uni-magdeburg.de}
\affiliation{Institute for Theoretical Physics, 
Otto-von-Guericke-University, PF 4120, D-39016 Magdeburg, Germany}
\author{H. B\"ottger}
\affiliation{Institute for Theoretical Physics, 
Otto-von-Guericke-University, PF 4120, D-39016 Magdeburg, Germany}
\author{V. V. Bryksin}
\affiliation{A. F. Ioffe Physical-Technical Institute, Politeknicheskaya 26,
19526 St.\ Petersburg, Russia}
\date{\today}

\begin{abstract}
A two dimensional hopping system with Rashba spin-orbit interaction is
considered. Our main interest is concerned with the evolution of the spin
degree of freedom of the electrons. We derive the rate equations governing
the evolution of the charge density and spin polarization of this system in
the Markovian limit in one-particle approximation. If only two-site hopping
events are taken into account, the evolution of the charge density and of
the spin polarization is found to be decoupled. A critical electric field is
found, above which oscillations are superimposed on the temporal decay of
the total polarization. A coupling between charge density and spin
polarization occurs on the level of three-site hopping events. The coupling
terms are identified as the anomalous Hall effect and the recently proposed
spin Hall effect. Thus, an unpolarized charge current through a sheet of
finite width leads to a transversal spin accumulation in our model system.
\end{abstract}

\maketitle

\section{Introduction}

The emerging field of spintronics tries to put the spin degree of
freedom of electrons to use in a fashion akin to electronics, where the
charge of the electrons is utilized.\cite{Prinz98,WolfEtal01} 
One of the advantages
of spintronics is, that spatial inhomogeneities of the spin distribution
are not burdened with an energetic penalty, quite in contrast to
inhomogeneous charge distributions. On the other hand, spin is not
conserved, whereas charge is. This is a disadvantage of spintronics,
which means that spin polarization normally decays with time.
Of special interest are techniques which influence the spin
by purely electrical means, without utilizing magnetic fields or magnetic
materials. Spin-orbit interaction (SOI) provides such a mechanism.

In recent years much
has been done in this field for itinerant electrons, favoring very clean
samples of high mobility.\cite{Martin03} This investigation, on the
other hand, studies the behavior of spins in a hopping system. Here,
mobilities are very low, either through disorder (Anderson localization), or
through strong electron-phonon interaction (polaron formation). Transport is
thermally activated and incoherent. In the absence of effective spin
scattering mechanisms, the low mobility might lead to long spin coherence
times for a given spin coherence length, 
which is one motivation for this study, apart from
the intrinsic scientific interest of the question of spin behavior in
hopping systems. 

Previous studies of SOI in hopping systems centered on the influence of the
spin dynamics on charge transport (e.g.\ 
magneto-conductance\cite{MovagharSchweitzer77,
MovagharSchweitzer78,Osaka79,ShapirOvadyahu89,
ShahbazyanRaikh94,Lin98a,Lin98b}), not on the spin dynamics
itself. Furthermore, the effective spin coupling between hopping sites has
normally been taken to be random,\cite{AsadaSlevinOhtsuki02,Lin98a,Lin98b,
MedinaKardar92,MedinaKardar91,MeirEtal91}
which is appropriate for spin scattering
on charged impurities, but not for SOI due to intrinsic or external electric
fields which are constant over length scales of several hopping lengths. But
such a mechanism would be required if one has the intent of affecting the
spins in a systematic way, i.e. realize spintronics.

To be specific, we consider a system where the spin dynamics is solely
determined by spin-orbit interaction through the 
Rashba-mechanism\cite{Rashba60,BychkovRashba84,MolenkampSchmidtBauer01,
PareekBruno02} and no
other spin scattering occurs (e.g.\ no magnetic impurities or spin
scattering on charged impurities). The electronic
system is two-dimensional (2D) and --- in order to calculate explicit
expressions --- assumed to be ordered (small polaronic system). The disordered
system is expected to behave qualitatively similar, but its investigation
must be delegated to future publications.

Two prominent examples of the interplay between charge and spin transport
are the anomalous Hall effect and the spin Hall effect: The observation that
a spin polarized charge current leads to a transversal Hall voltage, even
without an external magnetic field, is called the anomalous Hall effect (see
e.g.\ Ref.\
\onlinecite{CrepiouxBruno01} and references therein). 
The inverse effect, that, in materials showing the anomalous Hall effect,
an unpolarized charge current leads to a transversal
spin (but not charge!) current has been proposed by
Hirsch\cite{Hirsch99,Zhang00,HuGaoShen03} 
and is called the spin Hall effect. Both of these effects will in the
following be identified for our model system.

This paper is organized in the following way: Section \ref{secrsoi} is
concerned with the question of how to include the Rashba SOI into the hopping
formalism.
In Section \ref{secre} the rate equations governing the evolution
of the (1-particle) density matrix are derived. This is done by calculating
the diagrams of second and third order (two-site hopping and three-site
hopping) in the Konstantinov-Perel diagram technique in 1-particle
approximation and applying the Markovian limit. 
The rate equations for the density matrix are transformed into rate
equations for the particle density and the spin orientation, which is 
an equivalent representation, but much more lucid physically. 
It is found that two-site hopping does not introduce a coupling between
the equations for the particle (charge) density and the equations for the
spin polarization. Three-site hopping processes introduce such a coupling
and indeed lead to the anomalous Hall effect and the 
spin Hall effect for this hopping system. 

Sections \ref{sec2site} and \ref{sec3site} give solutions of the rate
equations for an ordered hopping system (i.e.\ a system of small polarons).
First, bulk properties (i.e.\ no boundaries) are considered. A short account
of the results of
Sec.\ \ref{sec2site} has been published in Ref.\
\onlinecite{DamkerBryksinBoettger04}.
Then, in Sec.\
\ref{sec3site}, a strip of finite width is considered, which introduces
boundary condition at the transversal edges. The stationary state with an
unpolarized charge current
shows spin accumulation at the boundaries, which is an expression of the
spin-Hall effect. 

The appendices primarily deal with mathematical details. Appendix
\ref{appmag} considers the magnetic field due to spin accumulation as a 
possible means of detection of spin accumulation.

\section{Rashba Spin-Orbit Interaction in Hopping Systems}
\label{secrsoi}
The first question which arises, is how to include the spin-orbit interaction
(SOI) into the formalism of hopping transport. The Rashba Hamiltonian reads
\begin{equation}
H=\frac{{\bf p}^2}{2m}-\frac{\hbar}{m}\boldsymbol{\sigma}\cdot({\bf K}
\times{\bf p})+V({\bf r}),
\label{hrashba}
\end{equation}
where we introduce the quantity ${\bf K}=m\alpha{\bf e}_z/\hbar^2$, which
(i) has the dimension of an inverse length, (ii) is perpendicular to the
two-dimensional plane (unit vector ${\bf e}_z$), and (iii) the length of
which signifies the Rashba SOI strength $\alpha$. The same expression Eq.\
(\ref{hrashba}) can be used for the generic SOI due to a spatially constant
field ${\bf E}$, in which case ${\bf K}=\frac{e}{4mc^2}{\bf E}$.
The symbol $\boldsymbol{\sigma}$ denotes the vector of Pauli spin
matrices.

Eq.\ (\ref{hrashba}) can be transformed to 
\begin{equation}
H=\frac{1}{2m}\left(
{\bf p}-\hbar{\boldsymbol{\sigma}}\times{\bf K}
\right)^2+V({\bf r}),
\label{hrashba2}
\end{equation}
while ignoring an irrelevant constant energy offset.
Thus, in principle, the Rashba
SOI can be dealt with as a SU(2) gauge potential. 
Since the gauge potential is a constant, this would be quite trivial, were
it not for the fact that SU(2) is non-Abelian. If the gauge were Abelian,
a wave function $\Psi({\bf r})$ would be studded with the factor
$\exp(i\boldsymbol{\sigma}\cdot({\bf K}\times{\bf r}))$ in order to include
SOI. Due to the non-Abelian nature, there are additional higher order
contributions in ${\bf K}$. Those can be neglected, provided $K$ is
sufficiently small.

The Hamiltonian which we consider reads
\begin{equation}
H=\sum_{m\sigma}\epsilon_m a_{m\sigma}^\dagger a_{m\sigma}
+\sum_{mm^\prime\sigma\sigma^\prime}
J^{\sigma^\prime\sigma}_{m^\prime m}
a_{m^\prime\sigma^\prime}^\dagger a_{m\sigma}
+H_{e-ph}+H_{ph},
\end{equation}
where $H_{e-ph}$ denotes the electron-phonon interaction and $H_{ph}$ is the
Hamiltonian of the phonon system. Here, $a^\dagger_{m\sigma}$
($a_{m\sigma}$) is the creation (annihilation) operator of an electron at
site $m$ in spin state $\sigma$, $\epsilon_m$ is the site energy, and 
$J^{\sigma^\prime\sigma}_{m^\prime m}$ is the resonance integral between the
states $m\sigma$ and $m^\prime\sigma^\prime$. Without SOI, the resonance
integral is diagonal in spin space and will be denoted as $J_{m^\prime m}$
in the following. Note, that the usual spin splitting due to Rashba 
SOI is not manifest
here, since Kramer's degeneracy demands, that the two spin states of a
localized eigenstate have to be degenerate.

The central question for the inclusion of SOI into the hopping Hamiltonian
now consists in determining the spin structure of $J_{m^\prime
m}^{\sigma^\prime\sigma}$. The four numbers belonging to each pair of sites
$m^\prime m$ are collected into a $2\times 2$-matrix in spin space
$\hat{J}_{m^\prime m}$.
 Clearly $\hat{J}_{m^\prime
m}\to J_{m^\prime m}\hat{I}_2$ for ${\bf K}\to{\bf 0}$, where $\hat{I}_2$
denotes the two-dimensional unit matrix. Since the  
Hamiltonian has furthermore to be Hermitic and invariant under time-reversal 
symmetry, the spin structure is given by an SU(2) 
matrix,\cite{FriedelLenglartLeman64,ZanonPichard88,
MeirGefenEntinWohlman89,Ando89,Lin98a,Lin98b,
AsadaSlevinOhtsuki02,PareekBruno02} 
$\hat{J}_{m^\prime
m}=\exp(-{\rm i}\boldsymbol{\sigma}\cdot{\bf A}_{m^\prime m})
J_{m^\prime m}$, where all components of the vector ${\bf A}$ have to be
real numbers and the condition
${\bf A}_{m^\prime m}=-{\bf A}_{mm^\prime}$ must be obeyed.
The arguments after Eq.\ (\ref{hrashba2}) show that the term linear in 
${\bf K}$ reads
$-{\rm i}\boldsymbol{\sigma}\cdot({\bf K}\times{\bf R}_{m^\prime m})$, 
where ${\bf R}_{m^\prime m}={\bf R}_{m^\prime}-{\bf R}_m$ is the distance
vector between the sites, and ${\bf R}_m$ is the coordinate vector of site
$m$. Thus,
\begin{equation}
\hat{J}_{m^\prime m}=e^{-{\rm i}\boldsymbol{\sigma}\cdot({\bf K}\times {\bf
R}_{m^\prime m})}J_{m^\prime m},
\end{equation}
For this expression for $\hat{J}$ to be valid, ${\bf K}$ has to be
sufficiently small, as explained in the paragraph after Eq.\
(\ref{hrashba2}). 
In this case, this means, that the condition
${\bf K}\times {\bf R}_{m^\prime m}\ll 1$ must be valid for the distance
vectors ${\bf R}_{m^\prime m}$, which are relevant to the case at hand
(e.g.\ the typical hopping length or the lattice constant).
The same expression, but applied to the Green's function of a hopping system,
has been derived in Ref.\ \onlinecite{ShahbazyanRaikh94}.

In some sense, this is equivalent to the Holstein
transformation,\cite{Holstein61} where the
influence of a magnetic field on a hopping system is taken into account
through a phase factor, and where higher order (in the magnetic field)
corrections are neglected.

We are now prepared to derive rate equations governing the hopping system
with the inclusion of Rashba SOI in the next section.

\section{Rate Equations}
\label{secre}
We apply the Konstantinov-Perel diagram technique in order to obtain rate
equations for the density matrix. The off-diagonal elements of the 
(1-electron) density matrix $\langle a^\dagger_{m^\prime\sigma^\prime}
a_{m\sigma}\rangle$ can usually be neglected in a hopping system (otherwise,
the transport mechanism would not be hopping). Here, we have to keep in
mind, that the correlations between different spin states on the same site
contain information regarding the (expectation value of the) spin
orientation. Furthermore, the model which we consider, aims at systems where
the spin degrees of freedom do not de-cohere (lose their phase memory) while
the electron stays on a site. Thus, we must retain the off-diagonal
elements in spin space {\emph{on the same site} (i.e.\ diagonal in the site
index) of the density matrix. In
this way, by including the SOI 
the occupation probability $\rho_m=\langle a^\dagger_m a_m\rangle$ becomes
a $2\times 2$-matrix $\hat{\rho}_m\vert_{\sigma^\prime\sigma}=\langle
a^\dagger_{m\sigma^\prime}a_{m\sigma}\rangle$ in spin space.

In order to adapt the Konstantinov-Perel diagram technique to the SOI case,
each site index can be thought of as additionally containing the spin index.
The summations run over all sites and also the two spin states. The
electron-phonon interaction is unaffected by the spin state. Thus, the
extension of the formalism reduces to (i) studding each site index with a
spin index (and extending the corresponding sums) and (ii) allowing
expectation values which are non-diagonal in spin space. The number of terms
in a given diagram proliferates quite rapidly with the order of the diagram
and makes its computation tedious. The calculations are greatly simplified,
if one replaces the explicit summation over the spin indices by matrix
multiplication in spin space. This can be achieved by collecting the
interaction matrix elements
$J$ (which depend on two site and two spin indices) into appropriate
matrices in spin space $\hat{J}_{m^\prime m}$ (which only depend on two site
indices) and doing likewise with the density matrix elements. 

The second order diagrams (two-site hopping) 
yield in one-particle approximation and in the Markovian limit the rate
equations
\begin{equation}
\frac{d}{dt}\hat{\rho}_m=\sum_{m_1}\left\{
e^{-{\rm i}\boldsymbol{\sigma}\cdot({\bf K}\times{\bf R}_{m_1m})}
\hat{\rho}_{m_1}
e^{{\rm i}\boldsymbol{\sigma}\cdot({\bf K}\times{\bf R}_{m_1m})}
W_{m_1m}
-\hat{\rho}_m W_{mm_1}
\right\},
\label{rate1}
\end{equation}
where the transition rates $W_{m^\prime m}$ are the same as are
obtained without SOI.

The physical meaning of Eq.\ (\ref{rate1}) is easier to assess, if
the following transformation is applied:
Representing the $2\times2$-density matrix at site $m$ by
$\hat{\rho}_m=\frac{1}{2}\rho_m\hat{I}_2
+\frac{1}{2}\boldsymbol{\rho}_m\cdot\hat{\boldsymbol{\sigma}}$, where
$\hat{I}_2$ is the two-dimensional unit matrix, the occupation probability
$\rho_m={\rm Tr}(\hat{\rho}_m)$ and the (expectation value of
the) spin orientation
$\boldsymbol{\rho}_m={\rm Tr}(\boldsymbol{\sigma}\hat{\rho}_m)$ are
introduced.
Please note, that in the following we will use the wording 
``spin orientation'' without emphasizing each time, that its expectation
value is meant. Of course, a spin-$1/2$ particle does not have a classical
spin orientation. 

The factors (matrices in spin space) surrounding
$\hat{\rho}_{m_1}$ in Eq.\ (\ref{rate1}) are inverses of each other. Thus,
by taking the trace (in spin space) of Eq.\ (\ref{rate1}) these factors
disappear and one obtains
\begin{equation}
\frac{d}{dt}\rho_m=\sum_{m_1}\left\{\rho_{m_1}W_{m_1m}-\rho_mW_{mm_1}
\right\}.
\label{raterho2m}
\end{equation}
On the other hand, multiplying Eq.\ (\ref{rate1}) by $\boldsymbol{\sigma}$
and then taking the trace, one obtains the rate equation for the spin
orientation (see Appendix \ref{appsigma})
\begin{equation}
\frac{d}{dt}\boldsymbol{\rho}_m=\sum_{m_1}\left\{
\hat{D}_{m_1m}\cdot\boldsymbol{\rho}_{m_1} W_{m_1m}
-\boldsymbol{\rho}_m W_{mm_1}
\right\},
\label{ratevecrho2m}
\end{equation}
where the quantity $\hat{D}$ is a
$3\times 3$-matrix describing a rotation about the axis 
${\bf A}_{m_1m}={\bf K}\times{\bf R}_{m_1m}$.
The Eqs.\ (\ref{raterho2m}) and (\ref{ratevecrho2m}) are the sought for rate
equations which replace Eq.\ (\ref{rate1}).

One can see, that in two-site approximation the equations for the occupation
number and the spin orientation decouple.  
Note, that the terms of
order ${\bf K}^2$ are retained in Eq.\ (\ref{ratevecrho2m}),
in addition to the terms of order ${\bf K}$, since the
third-order terms, to be derived next, give contributions only to second
order.
The behavior of an ordered (polaronic) hopping system obeying Eqs.\
(\ref{raterho2m}) and (\ref{ratevecrho2m}) is considered in Section
\ref{sec2site}.

Next, the three-site probabilities (third order diagrams) are calculated. 
These contributions yield higher order corrections to the two-site
probabilities, as well as some terms, which go beyond the physics of the
two-site expressions. Only the latter type of third order terms will be
retained in the following calculations. The products of spin matrices
occurring in the third order terms are calculated in Appendix \ref{appsigma}.
The rate
equations up to terms of order ${\bf K}^2$ and neglecting superfluous terms
as mentioned above read
\begin{eqnarray}
\frac{d}{dt}\rho_m&=&\sum_{m_1}\left\{W_{m_1m}\rho_{m_1}-W_{mm_1}\rho_m
\right\}\nonumber\\
&&{}+\sum_{m_1m_2}{\bf K}\cdot\left({\bf R}_{mm_1}\times{\bf R}_{mm_2}\right)
\circ{\bf K}\cdot\left\{W^R_{m_1m_2m}\boldsymbol{\rho}_{m_1}-W^R_{mm_1m_2}
\boldsymbol{\rho}_m
\right\}\label{ddtrhom}\\
\frac{d}{dt}\boldsymbol{\rho}_m&=&
\sum_{m_1}\left\{
\hat{D}_{m_1m}\cdot\boldsymbol{\rho}_{m_1} W_{m_1m}
-\boldsymbol{\rho}_m W_{mm_1}\right\}\nonumber\\
&&{}+\sum_{m_1m_2}{\bf K}\cdot\left({\bf R}_{mm_1}\times{\bf R}_{mm_2}\right)
\circ{\bf K}\left\{W^R_{m_1m_2m}\rho_{m_1}-W^R_{mm_1m_2}
\rho_m
\right\}\nonumber\\
&&{}-\sum_{m_1m_2}{\bf K}\cdot\left({\bf R}_{mm_1}\times{\bf R}_{mm_2}\right)
\circ{\bf K}\times\left\{W^I_{m_1m_2m}\boldsymbol{\rho}_{m_1}-W^I_{mm_1m_2}
\boldsymbol{\rho}_m
\right\}.
\label{ddtvecrhom}
\end{eqnarray}
Eqs.\ (\ref{ddtrhom}) and (\ref{ddtvecrhom}) are the generic hopping rate
equations for the model considered here (only Rashba SOI affects spin). The
transition rates $W_{m^\prime m}$, $W^R_{m_1m_2m}$, and $W^I_{m_1m_2m}$ are
independent of spin and are determined by the type of hopping considered
(e.g.\ small polaron hopping or hopping between impurities). In the
remaining part of the paper, we focus on an ordered (polaronic) system, but
Eqs.\ (\ref{ddtrhom}) and (\ref{ddtvecrhom}) would also be the starting
point for an investigation of a disordered system. Summing Eq.\
(\ref{ddtrhom}) over all sites $m$ yields $\frac{d}{dt}\sum_m\rho_m=0$,
thus, the derived rate equation for the particle density obeys particle (and
charge) conservation.

Whereas the second order transition rates $W_{m_1m}$ are those of a system
without SOI, the third order transition rates $W_{mm_1m_2}^{R\text{ or }I}$
differ from the corresponding quantities without SOI. Without SOI, the third
order transition rates (let us call them $W^{(3)}$ summarily) are important
for the Hall effect. The $W^{(3)}$ are divided into parts which are symmetric
in the magnetic field and parts which are anti-symmetric. Only the latter
are important for the Hall effect, thus the symmetric parts are normally
neglected. Here, since the magnetic field is zero, the anti-symmetric part
vanishes. Thus, we have to calculate the usually neglected symmetric parts
of $W^{(3)}$. Furthermore, both, real ($W^R$) and imaginary ($W^I$) part of
$W^{(3)}$ find entry in the formalism, whereas only the real part of the
second order $W$ occurs. Nevertheless, $W^R$ is connected with the Hall
mobility of a hopping system. The calculations are shown in Appendix
\ref{appw3}.

If the system is ordered (i.e.\ polaronic), it is advantageous to
transform the equations into wave-vector space. Note that, since the spatial
vectors are two-dimensional, the wave-vectors ${\bf q}$ will also be
two-dimensional. This transformation leads to the rate equations
\begin{equation}
\frac{d}{dt}\rho({\bf q})=\left[W({\bf q})-W({\bf 0})\right]
\rho({\bf q})+\left[{\bf W}^R({\bf q})-{\bf W}^R({\bf 0})\right]\cdot
\boldsymbol{\rho}({\bf q})
\end{equation}
and
\begin{eqnarray}
\frac{d}{dt}\boldsymbol{\rho}({\bf q})&=&
(W({\bf q})-W({\bf 0}))\boldsymbol{\rho}({\bf q})
-2\hat{W}({\bf q})\cdot\boldsymbol{\rho}({\bf q})
-2({\bf K}\times{\bf W}({\bf q}))\times\boldsymbol{\rho}({\bf q})
\nonumber\\
&&+({\bf W}^R({\bf q})-{\bf W}^R({\bf 0}))\rho({\bf q})
-({\bf W}^I({\bf q})-{\bf W}^I({\bf 0}))\times\boldsymbol{\rho}({\bf q})
\end{eqnarray}
Here, 
\begin{equation}
{\bf W}({\bf q})=\frac{1}{\rm i}\frac{\partial}{\partial{\bf q}}
W({\bf q})
\end{equation}
\begin{equation}
\hat{W}({\bf q})=-\left(
{\bf K}^2\frac{\partial}{\partial{\bf q}}\circ\frac{\partial} {\partial{\bf
q}}+{\bf K}\circ{\bf K}\frac{\partial}{\partial{\bf q}}\cdot
\frac{\partial}{\partial{\bf q}}
\right)W({\bf q})
\end{equation}
\begin{equation}
{\bf W}^{RI}({\bf q})={\bf K}\circ{\bf K}\cdot\left.\left(
\frac{\partial}{\partial{\bf q}}\times\frac{\partial}
{\partial{\bf q}_1}
\right)W^{RI}({\bf q}{\bf q}_1)\right|_{{\bf q}_1=0}
\label{vecwriq}
\end{equation}
where the last equation is to be understood as applying to the superscripts
$R$ and $I$, respectively.

The basic transition rates are
\begin{equation}
W({\bf q})=\sum_m e^{{\rm i}{\bf q}\cdot{\bf R}_m}W_{0m}
\end{equation}
\begin{equation}
W^{RI}({\bf q}{\bf q}_1)=\sum_{m_1m}e^{{\rm i}{\bf q}\cdot{\bf R}_m}
e^{{\rm i}{\bf q}_1\cdot{\bf R}_{m_1}}W^{RI}_{0m_1m}
\end{equation}
In an isotropic system,
taking the long wavelength limit, they read
\begin{equation}
W({\bf q})=W({\bf 0})-D{\bf q}^2-{\rm i}\mu{\bf q}\cdot{\bf E}
\end{equation}
and
\begin{equation}
{\bf W}^{RI}({\bf q})={\rm i}{\bf K}\circ{\bf q}\cdot({\bf E}\times{\bf K})
W^{RI},
\end{equation}
the latter being calculated for a triangular lattice 
(see Appendix \ref{appw3}). Here, $D$ is the diffusion constant, $\mu$ the
mobility, and the quantities $W^R$ and $W^I$ are constants defined through
Eq.\ (\ref{wrofq}).

The rate equations thus take the shape
\begin{equation}
\frac{d}{dt}\rho({\bf q})=-{\rm i}{\bf q}\cdot{\bf j}({\bf q})
\label{continuitycharge}
\end{equation}
with the particle current density (proportional to the charge current density
$e{\bf j}$)
\begin{equation}
{\bf j}({\bf q})=\mu{\bf E}\rho({\bf q})-{\rm i}D{\bf q}\rho({\bf q})
+W^R{\bf K}\times{\bf E}\circ{\bf K}\cdot\boldsymbol{\rho}({\bf q})
\label{chargecurrent}
\end{equation}
and
\begin{equation}
\frac{d}{dt}\boldsymbol{\rho}({\bf q})=-{\rm i}{\bf q}\cdot\hat{J}({\bf
q})+{\bf Q}({\bf q})
\label{continuityspin}
\end{equation}
with the (tensorial) spin current density
\begin{eqnarray}
\hat{J}({\bf q})&=&\mu{\bf E}\circ\boldsymbol{\rho}({\bf q})
-{\rm i}D{\bf q}\circ\boldsymbol{\rho}({\bf q})
+4D\hat{I}_3{\bf K}\cdot\boldsymbol{\rho}({\bf q})
-4D\boldsymbol{\rho}({\bf q})\circ{\bf K}
\nonumber\\
&&+W^R{\bf K}\times{\bf E}
\circ{\bf K}\rho({\bf q})
-W^I{\bf K}\times{\bf E}
\circ{\bf K}\times\boldsymbol{\rho}({\bf q})
\label{spincurrent}
\end{eqnarray}
and the spin source density 
\begin{equation}
{\bf Q}({\bf q})=-4D{\bf K}^2\boldsymbol{\rho}({\bf q})
-4D{\bf K}\circ{\bf K}\cdot\boldsymbol{\rho}({\bf q})
+2\mu({\bf K}\times{\bf E})\times\boldsymbol{\rho}({\bf q})
\label{spinsource}
\end{equation}

Eq.\ (\ref{continuitycharge}) is the continuity equation for the
particle (charge) density. Eq.\ (\ref{continuityspin}) contains a source
term in addition to the divergence of the current, since there is no
conservation law for spin. The first two terms contributing to the charge
current (\ref{chargecurrent}) are the drift term in the electrical field and
the diffusion. The third term only occurs, when there is a finite
$z$-component of the polarization (i.e.\ a magnetization $M_z$). The
resulting current is directed perpendicular to the electric field. Thus,
this term corresponds to the anomalous Hall effect. The corresponding
mobility is seen to be $\mu_{yx}=W^RK^2$. The quantity $\mu_{yx}/\mu$ is
related to the Hall mobility (see Appendix \ref{appw3}).

The spin current Eq.\ (\ref{spincurrent}) mainly 
consists of the drift term and some diffusion terms.
But the most interesting contribution is the fifth
term on the right-hand side, since it is also present for vanishing spin
polarization $\boldsymbol{\rho}={\bf 0}$. This term describes a current of
$z$-spins into the direction perpendicular to the electric field 
(${\bf K}\times{\bf E}$). Thus, it is an expression of the spin Hall effect. 
The co-efficient $W^R$ is the same, as the one responsible for 
the anomalous Hall effect.

In order to simplify notation, dimensionless quantities are introduced:
dimensionless space ${\bf x}=K{\bf r}$, wave vector $\boldsymbol{\xi}={\bf
q}/K$, time $\tau=DK^2t$ and electric field
$\boldsymbol{\epsilon}={\mu{\bf E}}/({DK})$. 
The transformed third order
probabilities are denoted as
$\epsilon_R=\frac{KE}{D}W^R$ and 
$\epsilon_I=\frac{KE}{D}W^I$. Note, that $\epsilon_R$ can be expressed as 
$\mu_{yx}\epsilon/\mu$.
Thus, $\epsilon_R$ is an analog of the Hall mobility multiplied by the
electrical field.
Furthermore, the co-ordinate system is fixed
such, that ${\bf e}_z\propto{\bf K}$, ${\bf e}_x\propto{\bf E}$, and ${\bf
e}_y\propto{\bf K}\times{\bf E}$. Then ${\bf K}=K{\bf e}_z$ and ${\bf
E}=E{\bf e}_x$.
This transforms the rate equations (\ref{continuitycharge}) and
(\ref{continuityspin}) to
\begin{equation}
\frac{d}{d\tau}\rho(\boldsymbol{\xi},\tau)
=-(\boldsymbol{\xi}^2+{\rm i}\xi_x\epsilon)\rho(\boldsymbol{\xi},\tau)
-{\rm i}\epsilon_R\xi_y\rho_z(\boldsymbol{\xi},\tau)
\label{ddtrhoxi}
\end{equation}
and
\begin{eqnarray}
\frac{d}{d\tau}\boldsymbol{\rho}(\boldsymbol{\xi},\tau)&=&
-(\boldsymbol{\xi}^2+{\rm i}\xi_x\epsilon)\boldsymbol{\rho}
-4{\rm i}({\bf e}_z\times\boldsymbol{\xi})
\times\boldsymbol{\rho}
-{\rm i}\epsilon_R\xi_y{\bf e}_z\rho+{\rm i}\epsilon_I\xi_y
{\bf e}_z\times\boldsymbol{\rho}
\nonumber\\
&&-4\boldsymbol{\rho}
-4{\bf e}_z\rho_z
+2\epsilon{\bf e}_y\times\boldsymbol{\rho}
\label{ddtvecrhoxi}
\end{eqnarray}

The three terms on the right-hand side of Eq.\ (\ref{ddtrhoxi}) are the
diffusion contribution, the drift in the electric field and the anomalous
Hall effect. The first terms on the right-hand side of 
Eq.\ (\ref{ddtvecrhoxi}) are: diffusion, drift (Ohmic current), a precession
term for diffusing spins, the spin Hall effect, and a precession of the spin
about the
$z$-axis. The last three terms contribute to the spin decay. The first two
of those can be expanded as $-4{\bf e}_x\rho_x-4{\bf e}_y\rho_y-8{\bf
e}_z\rho_z$, thus the decay rate for the $z$-component is twice as large
compared with the in-plane components.

\section{Bulk System}
\label{sec2site}

By setting the wave vector $\boldsymbol{\xi}$ to zero, the evolution
equations for the quantities integrated over the whole system (total charge
and total spin polarization, e.g.\  
$\rho(\boldsymbol{\xi}={\bf 0},\tau)=\int d^2{\bf x}\rho({\bf x},\tau)$) are
obtained 
\begin{eqnarray} 
\frac{d}{d\tau}\rho({\bf 0},\tau)&=&0\label{rhotot}\\ 
\frac{d}{d\tau}\boldsymbol{\rho}({\bf 0},\tau)
&=&-4\boldsymbol{\rho}-4{\bf e}_z\rho_z
+2\epsilon{\bf e}_y\times\boldsymbol{\rho}
.\label{vecrhotot}
\end{eqnarray}
Eq.\ (\ref{rhotot}) is an expression of charge conservation. 
It is to be seen, that the evolution of the space integrated quantities does
not depend on the third-order parameters $\epsilon_R$ and $\epsilon_I$.

The three components of Eq.\ (\ref{vecrhotot}) read
\begin{subequations}
\begin{eqnarray}
\frac{d}{d\tau}\rho_x&=&-4\rho_x+2\epsilon\rho_z
\label{ddtrhoxtot}\\
\frac{d}{d\tau}\rho_y&=&-4\rho_y\\
\frac{d}{d\tau}\rho_z&=&-8\rho_z-2\epsilon\rho_x,
\label{ddtrhoztot}
\end{eqnarray}
\label{drhodttot}
\end{subequations}
the solution of which is
\begin{subequations}
\begin{eqnarray}
\rho_x(\tau)&=&e^{-6\tau}\left[
\rho_x(0)\cosh(2\tau\sqrt{1-\epsilon^2})
+\frac{\rho_x(0)+\epsilon\rho_z(0)}{\sqrt{1-\epsilon^2}}
\sinh(2\tau\sqrt{1-\epsilon^2})
\right]\\
\rho_y(\tau)&=&e^{-4\tau}\rho_y(0)\\
\rho_z(\tau)&=&e^{-6\tau}\left[
\rho_z(0)\cosh(2\tau\sqrt{1-\epsilon^2})
-\frac{\epsilon\rho_x(0)+\rho_z(0)}{\sqrt{1-\epsilon^2}}
\sinh(2\tau\sqrt{1-\epsilon^2})
\right].
\end{eqnarray}
\label{vecrhotau}
\end{subequations}
$\rho_x(0)$, $\rho_y(0)$, and $\rho_z(0)$ are the initial conditions.
Note, that there is a critical electrical field strength $\epsilon_c=1$. In
smaller fields the total spin polarization decays exponentially with time,
but in larger fields, it additionally rotates with time.
This critical electric field $E_c=DK/\mu$ can also be expressed as
$E_c=k_BTK/e$, provided the Einstein relation between $D$ and $\mu$ is valid.
Furthermore, without electric field, the $z$-component of
the total spin polarization decays twice as fast as the other two
components.\cite{DamkerBryksinBoettger04,Bleibaum04}

Taking for definiteness the initial condition $\boldsymbol{\rho}(0)={\bf
e}_z$, one obtains
\begin{subequations}
\begin{eqnarray}
\rho_x(\tau)&=&e^{-6\tau}
\frac{\epsilon}{\sqrt{1-\epsilon^2}}
\sinh(2\tau\sqrt{1-\epsilon^2})\\
\rho_y(\tau)&=&0\\
\rho_z(\tau)&=&e^{-6\tau}\left[
\cosh(2\tau\sqrt{1-\epsilon^2})
-\frac{1}{\sqrt{1-\epsilon^2}}
\sinh(2\tau\sqrt{1-\epsilon^2})
\right]
\end{eqnarray}
\end{subequations}
for $\epsilon<1$, whereas for $\epsilon>1$ the hyperbolic functions become
trigonometric functions. Expressed differently, for $\epsilon>1$ 
the eigen-values of the set of equations (\ref{drhodttot})
become complex, which means that its eigen-modes become
oscillatory. Thus,
\begin{subequations}
\begin{eqnarray}
\rho_x(\tau)&=&e^{-6\tau}
\frac{\epsilon}{\sqrt{\epsilon^2-1}}
\sin(2\tau\sqrt{\epsilon^2-1})\\
\rho_y(\tau)&=&0\\
\rho_z(\tau)&=&e^{-6\tau}
\frac{\epsilon}{\sqrt{\epsilon^2-1}}
\cos(2\tau\sqrt{\epsilon^2-1}+\phi),
\end{eqnarray}
\end{subequations}
with $\sin\phi=1/\epsilon$.
Note, that the frequency of the polarization oscillation
$2\sqrt{\epsilon^2-1}$ increases with the electric field $\epsilon$. Apart
from the decay factor $e^{-6\tau}$, the polarization
$\boldsymbol{\rho}(\tau)$ describes a slanted ellipse in the $x$-$z$-plane,
i.e.\ the plane spanned by the Rashba field ${\bf K}$ 
and the electric field ${\bf E}$.
The co-ordinates
$x=\rho_xe^{6\tau}\sqrt{\epsilon^2-1}/\epsilon$ and 
$z=\rho_ze^{6\tau}\sqrt{\epsilon^2-1}/\epsilon$ determine the ellipse
$x^2+2xz\sin\phi+z^2=\cos^2\phi$.
The inclination of this ellipse (angle between $z$-axis and the main
diagonal) increases with electric field. For $\epsilon\to 1$ the angle
$\phi\to\pi/2$, i.e.\ the ellipse degenerates to a line with an inclination
of 45$^\circ$ to the perpendicular direction. In the limit of strong
electric field $\epsilon\to\infty$, the angle $\phi$ vanishes, and the
ellipse becomes a circle.
In order to observe these
oscillations, the corresponding frequency should be larger or of the order
of the inverse decay time constant, which leads in this case to the
approximate condition $\epsilon\agt 10$.

Let us consider next the case of an inhomogeneous initial condition. We take
a $z$-spin starting at the origin
$\boldsymbol{\rho}({\bf x},\tau=0)=\delta({\bf x}){\bf e}_z$.
The total spin polarization is still described by Eqs.\ (\ref{vecrhotau}). 
In order to obtain the local spin orientation, one has to solve Eqs.\
(\ref{ddtrhoxi}) and (\ref{ddtvecrhoxi}). This is done by solving these
equations in Laplace-space and the following back-transformation from
Laplace-space $s$ to time $\tau$ and wave-vector space $\boldsymbol{\xi}$ 
to ${\bf x}$.

The asymptotic behavior of the local spin polarization 
for large times $\tau\to\infty$ and
small distances $\vert{\bf x}\vert\ll 2\sqrt{\tau}$ (Large distances will 
not be relevant here, since they are strongly suppressed by the 
``diffusion factor'' $\exp(-{\bf x}^2/4\tau)$.) is
\begin{equation}
\rho_z({\bf x},\tau)=\exp\left\{
-\frac{({\bf x}+\boldsymbol{\epsilon}\tau)^2}{4\tau}
-\frac{7}{4}\tau
\right\}
\frac{3}{8\sqrt{\pi\tau}}
\textrm{J}_0\left(\frac{\sqrt{15}}{2}\vert {\bf x}\vert\right)
\label{rhozxtau}
\end{equation}
Note, that in this approximation (long times and short distances),
the sign of the polarization --- determined by the argument of the Bessel
function $J_0$ --- only depends on the distance to the
origin, and is furthermore not time dependent. 
This spatial oscillation is due to the precession terms in the rate
equations. The exponential factor exhibits a decay term with (dimensionless)
time constant $4/7$ and a drift-diffusion term with 
drift velocity $-\boldsymbol{\epsilon}$ and
diffusion constant $1$.

\section{Strip of finite width}
\label{sec3site}
The distinguishing feature of a strip of finite width is, that
there is no current (neither charge, nor spin) 
across the transversal boundaries. Thus, the spin-Hall current must be
compensated by a diffusion current, i.e.\ a finite spin density can be
expected to occur near the boundary.
Indeed, the following will show, that
a transversal spin accumulation will occur in this situation. For a broad
strip, the non-zero spin density is mainly constricted to near the boundaries.

The strip is oriented, so that the $x$-direction is along the
length of the strip (the direction of the current flow), 
the $y$-direction is along the width of the strip and the $z$-direction is
perpendicular to the 2D strip (see Fig.\ \ref{figstrip}). The strip extends
from $y=-b$ to $y=b$, thus, its width is $2b$.
Going from wave-vector space back to ${\bf x}$-space and 
assuming that $\rho({\bf x})$ and 
$\boldsymbol{\rho}({\bf x})=(\rho_x,\rho_y,\rho_z)$ depend only on
the $y$-coordinate, the following set of differential equations determine
the stationary state
\begin{subequations}
\begin{eqnarray}
\rho^{\prime\prime}+\epsilon_R\rho_z^\prime&=&0\label{stripfirst}\\
\rho_x^{\prime\prime}+\epsilon_I\rho_y^\prime+2\epsilon\rho_z-4\rho_x&=&0\\
\rho_y^{\prime\prime}-\epsilon_I\rho_x^\prime+4\rho_z^\prime-4\rho_y&=&0\\
\rho_z^{\prime\prime}+\epsilon_R\rho^\prime-4\rho_y^\prime-2\epsilon\rho_x
-8\rho_z&=&0
\end{eqnarray}
\end{subequations}
where the prime denotes the derivation with respect to $y$. 
The boundary conditions at $y=\pm b$ are obtained by equating the normal
component of the currents (\ref{chargecurrent}) and (\ref{spincurrent}) 
at the boundary to zero. This yields
\begin{subequations}
\begin{eqnarray}
(\rho^\prime+\epsilon_R\rho_z)\vert_{y=\pm b}&=&0\\
(\rho_x^\prime+\epsilon_I\rho_y)\vert_{y=\pm b}&=&0\\
(\rho_y^\prime-\epsilon_I\rho_x+4\rho_z)\vert_{y=\pm b}&=&0\\
(\rho_z^\prime+\epsilon_R\rho-4\rho_y)\vert_{y=\pm b}&=&0
\label{striplast}.
\end{eqnarray}
\end{subequations}

One has to keep in mind, that we want the solution to lowest order in $K$.
The zero-th order solution of Eqs.\ (\ref{stripfirst}) to (\ref{striplast})
is obtained by removing all terms at least linear in $K$, i.e.\ 
setting $\epsilon_I=0$ and $\epsilon_R=0$. This immediately yields the solution
$\rho(y)={\rm const.}$ and $\boldsymbol{\rho}(y)=0$.
This solution is inserted into the terms containing $\epsilon_R$ and
$\epsilon_I$,
which yields the equations for the first order solution
\begin{subequations}
\begin{eqnarray}
\rho^{\prime\prime}&=&0\\
\rho_x^{\prime\prime}+2\epsilon\rho_z-4\rho_x&=&0\\
\rho_y^{\prime\prime}+4\rho_z^\prime-4\rho_y&=&0\\
\rho_z^{\prime\prime}-4\rho_y^\prime-2\epsilon\rho_x-8\rho_z&=&0
\end{eqnarray}
\end{subequations}
with the boundary conditions
\begin{subequations}
\begin{eqnarray}
(\rho^\prime)\vert_{y=\pm b}&=&0\\
(\rho_x^\prime)\vert_{y=\pm b}&=&0\\
(\rho_y^\prime+4\rho_z)\vert_{y=\pm b}&=&0\\
(\rho_z^\prime-4\rho_y)\vert_{y=\pm b}&=&\epsilon_R\rho.
\label{boundinh}
\end{eqnarray}
\end{subequations}
The solution of this set of equations can easily be constructed, since the
only difference to the equations of zero-th order is the finite right-hand
side in the boundary condition (\ref{boundinh}).
Again, $\rho(y)={\rm const.}$, and the
3 vectorial $\rho$-components are weighted sums of functions $\sinh(\lambda
y)$ and $\cosh(\lambda y)$ where $\lambda$ takes on the three values:
$\lambda_1=2$, $\lambda_2=\sqrt{-2+2{\rm i}\sqrt{\epsilon^2+7}}=
\sqrt{\sqrt{8+\epsilon^2}-1}+{\rm i}\sqrt{\sqrt{8+\epsilon^2}+1}$ and
$\lambda_3=\sqrt{-2-2{\rm i}\sqrt{\epsilon^2+7}}=\lambda_2^*$.
Thus, to lowest order in ${\bf K}$,
\begin{subequations}
\begin{eqnarray}
\frac{\rho_x(y)}{\epsilon_R\rho}
&=&-\frac{16\epsilon}{\alpha}\lambda_2c_2s_3\sinh(2y)
+\frac{2\epsilon}{\alpha}\left(16c_1s_3+\frac{\epsilon^2}{2}\lambda_3
s_1c_3\right)\sinh(\lambda_2 y)+\text{c.c.}\\
\frac{\rho_y(y)}{\epsilon_R\rho}
&=&\frac{4\epsilon^2}{\alpha}\lambda_2c_2s_3\cosh(2y)
+\frac{4\lambda_2}{\alpha}\left(16c_1s_3+\frac{\epsilon^2}{2}\lambda_3s_1c_3
\right)\cosh(\lambda_2y)+\text{c.c.}\\
\frac{\rho_z(y)}{\epsilon_R\rho}
&=&-\frac{\lambda_2^2-4}{\alpha}\left(
16c_1s_3+\frac{\epsilon^2}{2}\lambda_3s_1c_3
\right)\sinh(\lambda_2y)+\text{c.c.}
\label{rhozofy}
\end{eqnarray}
\end{subequations}
with the abbreviations
\begin{equation}
\alpha=\lambda_2c_2\left[
\frac{\epsilon^2}{2}\vert\lambda_2\vert^2\lambda_2s_1c_3
+16(\lambda_2^2+12+\epsilon^2)c_1s_3
\right]-\text{c.c.}
\end{equation}
and
\begin{equation}
c_i=\cosh(\lambda_i b),\quad s_i=\sinh(\lambda_i b)\quad(i=1,2,3).
\end{equation}
One can see, that the electron density $\rho$ and the spin Hall
co-efficient $\epsilon_R$ only influence the amplitude of the polarization.
They do not affect the spatial dependence of $\boldsymbol{\rho}(y)$.
The $x$- and $z$-component of the polarization are anti-symmetric with
respect to a sign change of $y$, whereas the $y$-component is symmetric.

For a large electric field $\epsilon\gg 1$
\begin{subequations}
\begin{eqnarray}
\rho_x(y)&\approx&-{\rm sgn}(y)\frac{\epsilon_R\rho}{\sqrt{2\epsilon}}
\sin\left(\sqrt{\epsilon}(b-\vert y \vert)
+\frac{\pi}{4}\right)e^{-\sqrt{\epsilon}(b-\vert y\vert)}\\
\rho_y(y)&\approx&-\frac{\epsilon_R\rho}{\epsilon}
\sin\left(\sqrt{\epsilon}(b-\vert y\vert)\right)
e^{-\sqrt{\epsilon}(b-\vert y\vert)}\\
\rho_z(y)&\approx&-{\rm sgn}(y)\frac{\epsilon_R\rho}{\sqrt{2\epsilon}}
\cos\left(\sqrt{\epsilon}(b-\vert y \vert)+\frac{\pi}{4}\right)
e^{-\sqrt{\epsilon}(b-\vert y\vert)}
\label{rhozlargee}
\end{eqnarray}
\end{subequations}
The main effects of increasing the electric field $\epsilon$ consists in a 
sharper concentration of
the polarization near the boundary (due to the exponential term)
and an increase of the amplitude of the $x$- and $z$-component of the
polarization (because $\epsilon_R/\sqrt{\epsilon}=\sqrt{\epsilon}\mu_{yx}/\mu
\propto\sqrt{\epsilon}$).

For $b\gg 1$ (i.e.\ in natural units: width $\gg K^{-1}$)
the spin polarization is significant only near the
boundaries due to the hyperbolic functions. This suggests that there is a
``spin diffusion length'' governing the spatial extent of non-zero
$\boldsymbol{\rho}$ near the borders. This suggestion can be fleshed out in
the following way: We consider the half plane $y\ge 0$, set the electric
field $\epsilon$ to zero and apply the boundary condition
$\boldsymbol{\rho}(y=0)=\boldsymbol{\rho}_0$.
The differential equations for the stationary state 
(and homogeneous in $x$) are  
\begin{equation}
\rho_x^{\prime\prime}-4\rho_x=0,\quad
\rho_y^{\prime\prime}+4\rho_z^\prime-4\rho_y=0,\quad
\rho_z^{\prime\prime}-4\rho_y^\prime-8\rho_z=0,
\label{spindiffconst}
\end{equation}
which conceptually corresponds to an equation with the structure
$D\Delta\boldsymbol{\rho}-\boldsymbol{\rho}/\tau=0$, only that here the decay
rate $1/\tau$ depends on the direction (thus, it is a tensor), and there is
an additional ``torque'' term. Since in Eq.\ (\ref{spindiffconst}) all
co-efficients are numbers (in dimensionless units) the resulting ``spin
diffusion length'' $\sqrt{D\tau}$ is also a number of the order of one (in
the present case: $1/2$ for the $x$-component and
$1/\sqrt{\sqrt{8}-1}\approx 0.74$ for
the other two components, corresponding to the real parts of $\lambda_1$,
$\lambda_2$ and $\lambda_3$ at $\epsilon=0$). Thus, expressed in natural units, this length
is $K^{-1}$ times a constant of order one, i.e.\ this spin diffusion does
not introduce a new length scale in addition to the Rashba length scale
$K^{-1}$.

The spin current of a non-polarized electrical current
in an infinitely extended plane consists of $z$-spins being transported
perpendicularly to the electrical field ${\bf E}$ (see Eq.\
(\ref{spincurrent})). Here, this spin current is ``transformed'' into spin
accumulation by applying boundary conditions. Seen in this light, 
it is perhaps
somewhat surprising that the resulting spin polarization has finite $x$- and
$y$-components, in addition to the $z$-component. But this can be
explained by the following observation: The calculated spin polarization
$\boldsymbol{\rho}(y)$ is --- though stationary --- not static. It
represents an equilibrium between the ``spin supply'' by the spin Hall
effect on the one hand, and other contributions of the spin current as well
as the spin decay, on the other hand. Even though the ``supply'' only yields
$z$-spins, the tensorial nature of the current leads to spin precession, such
that in the equilibrium state all components occur.

Due to the spin polarization (magnetization) a magnetic field ${\bf H}_M$ is
generated, which may in principle allow to detect this polarization. In
order to actually measure this magnetic field, one has to take into account,
that there is also a magnetic field ${\bf H}_I$ due to the electric current
through the strip. These fields are calculated in Appendix \ref{appmag}.

In principle, the magnetic field ${\bf H}_M+{\bf H}_I$ interacts with the
spin degrees of freedom through the Zeeman energy and also applies a torque 
to the spins. These effects have been neglected in this paper.

The (in-)significance of the back-reaction of the magnetic field due to 
the spins on the spins can be estimated in the following way:
Using the magnetic susceptibility $\chi$ and the Bohr magneton $\mu_B$, the
spin polarization due to the back reaction can be
expressed as $\boldsymbol{\rho}_{\text{br}}=\chi {\bf H}/\mu_B
\approx\chi {\bf M}/\mu_B=\chi\boldsymbol{\rho}$. Thus, the influence of the
magnetic field on the spins can be neglected, provided the condition
$\chi\ll 1$ is fulfilled.

\section{Discussion}
\label{secdiss}
We have studied the spin dynamics of a hopping model where
Rashba SOI is the sole spin scattering mechanism. Rate equations have been
derived in the one-particle approximation and the Markovian limit, including
two-site and three-site hopping processes. Further calculations are
performed for the case of polaron hopping, where one can work in wave-vector
space.

As expected, an
initial spin polarization has been found to decay
exponentially with time,
if the electric field is low. But, there is a critical electric field, above
which the total spin polarization shows oscillations (rotations) overlaying
the decay. This can be understood as follows:

The rotation of a spin initially at a certain site $m_1$ during a single hop
to another site $m_2$ depends only on the difference vector 
${\bf R}_{m_2m_1}$. The spin rotates in the plane spanned by the vectors 
${\bf K}$ and ${\bf R}_{m_2m_1}$. Let us consider a spin starting at the 
origin and being orientated along the $z$-direction at $t=0$. 
The motion of the electron in the radial directions leads to 
concentric ``waves'' around the origin
of spins having the same $z$- and radial component.
But, due to the diffusive motion, the electron
does not only take the straight path radially outwards, but explores an
increasingly entangled trajectory, thereby randomizing the spin orientation.
The random motion only affects a decrease of the magnitude of the 
polarization (it neither
creates polarization, nor does it favor any specific orientation). Thus, the
preferred orientation given by the radial motion is preserved, but its
magnitude diminished.

An electric field does affect the charge transport, but in two-site hopping
approximation it does not affect
the elementary processes responsible for the transport of spin orientation.
Thus, in this case, 
the spin polarization is simply the product of particle density and
spin orientation, which has a constant magnitude itself (quite in contrast
to the situation with three-site hopping processes).
In this way, one can see, that the electric field does not change the
spatial distribution of the spin orientation (e.g.\ the sign of the
$z$-component), but only changes the temporal evolution of the spatial charge
distribution (the ``weight'' of the local spin orientation). This explains
the time independence of the argument of the Bessel function in Eq.\
(\ref{rhozxtau}).

On the other hand, in a very large electric field an electron follows
preferably a straight trajectory along the field (large drift component in
comparison to the diffusion). The
importance of diffusion about this trajectory is reduced, thereby also
reducing the randomization of the spin. This explains, why it is
possible for the electric field to become so strong, that the spatial
oscillations of the spin orientation survive the averaging over the whole
volume, so that they appear as temporal oscillations.

Without an electric field, the decay of an initial spin orientation is found
to behave anisotropic: The $z$-component (perpendicular to the plane) decays
twice as fast as the in-plane components.

The inclusion of three-site hopping terms leads to the occurrence of a
transport co-efficient ($\epsilon^R$ or $\mu_{yx}$) which provides a
coupling between charge and spin transport. A spin current produces a
transversal charge current (anomalous Hall effect), and a charge current
produces a transversal spin current (spin Hall effect). In a
two-dimensional strip of finite width, this transversal spin current
manifests itself as a spin accumulation at the boundaries of the strip. 

Finally, some remarks about the experimental relevance are in order here.
The Rashba SOI strength is widely given as being of the order
$\alpha\approx 10^{-9}$~eV~cm.\cite{ShahbazyanRaikh94, MolenkampSchmidtBauer01}
It is possible to change this value by applying a gate voltage to the
two-dimensional plane\cite{NittaEtal97}, where a variation by at least 
50 \% has been shown to be realizable. This might be conceptually 
favorable for the implementation of spintronics devices, where electric means
to control spins are sought.

Using the value of $\alpha$ mentioned above, one obtains 
$K\approx1/(76~\text{\AA})$, assuming
that the effective electron mass is equal to its elementary mass. 
Thus, the condition ${\bf K}\times{\bf R}_{m^\prime m}\ll 1$ for the
validity of our theory demands in this case, 
that the typical hopping length is shorter than 76~\AA. This
value is conveniently large, so that the range of polaronic 
nearest neighbor hopping is well within the validity range. 
But impurity hopping or variable range hopping are
also not out of the question. Furthermore,
for a smaller Rashba co-efficient, the permitted length
increases and it is furthermore modified by the effective electron mass.
If the Einstein relation between diffusion co-efficient and mobility is
assumed to be valid, the critical electric field $E_c=k_BTK/e$ 
for the oscillatory behavior of the total spin polarization is estimated 
to be about $10^4$~V/cm at a temperature of 100~K.

The calculations in Appendix \ref{appw3}
show a close connection between Hall mobility and the co-efficient for the
spin Hall effect. Thus, a large Hall mobility favors also the occurrence of
the spin Hall effect. We take
Ref.\ \onlinecite{RhyeeChoRi03} as an example. There, 
the conductivity and the 
Hall mobility of the hexaboride compounds Eu$_{1-x}$Ca$_x$B$_6$ are
measured, and the authors
conclude, that the transport mechanism is polaron hopping (in a
certain range of concentrations $x$ and for high temperatures). The reported
Hall mobility in the polaronic hopping regime is of the order of
100~cm$^2$/(Vs).
Using this value, 
Eq.\ (\ref{muyxvsmuh}) gives the estimate $\mu_{yx}/\mu\approx 0.23$, i.e.\
the transversal mobility for the spin Hall effect is about a fourth of the
longitudinal mobility. Taking Eq.\ (\ref{rhozofy}) into account, 
this gives a
relative spin polarization ($z$-component) at the boundary of a
two-dimensional strip $\vert\rho_z(b)/\rho\vert$ of about
5\% at the critical electric field strength $\epsilon=1$.

\begin{acknowledgments}
This work is in part supported by the DFG (Deutsche Forschungsgemeinschaft) 
under Grant No.\ 436 RUS 113/67/11-2.
\end{acknowledgments}

\appendix
\section{Calculation of the spin matrices}
\label{appsigma}

Applying the identity
$(\boldsymbol{\sigma}\cdot{\bf A})(\boldsymbol{\sigma}\cdot{\bf B})=({\bf
A}\cdot{\bf B})+{\rm i}\boldsymbol{\sigma}\cdot({\bf A}\times{\bf B})$,
valid for any vectors ${\bf A}$ and ${\bf B}$,
to the series expansion of the exponential function,
one easily obtains the relation
\begin{equation}
e^{{\rm i}\boldsymbol{\sigma}\cdot{\bf A}}=\cos\vert{\bf A}\vert+{\rm i}
\boldsymbol{\sigma}\cdot{\bf A}\frac{\sin\vert{\bf A}\vert}
{\vert{\bf A}\vert},
\label{eisigma}
\end{equation}
which is valid for any vector ${\bf A}$.

The quantity 
${\rm Tr}\left(\boldsymbol{\sigma}
e^{-{\rm i}\boldsymbol{\sigma}{\bf A}_{m_1m}}\hat{\rho}_{m_1}
e^{{\rm i}\boldsymbol{\sigma}{\bf A}_{m_1m}}\right)$ occurs
in the calculation of the contribution of two site diagrams to the rate
equations. It can be transformed to ${\rm Tr}\left(
e^{{\rm i}\boldsymbol{\sigma}{\bf A}_{m_1m}}\boldsymbol{\sigma}
e^{-{\rm i}\boldsymbol{\sigma}{\bf A}_{m_1m}}\hat{\rho}_{m_1}
\right)$. 
With the help of Eq.\ (\ref{eisigma}),
the spin factors in the trace can be calculated to be
\begin{equation}
e^{{\rm i}\boldsymbol{\sigma}{\bf A}_{m_1m}}\boldsymbol{\sigma}
e^{-{\rm i}\boldsymbol{\sigma}{\bf A}_{m_1m}}\approx
\hat{D}_{m_1m}\boldsymbol{\sigma},
\label{asigmaa}
\end{equation} 
where the quantity $\hat{D}$ is a
$3\times 3$-matrix describing a rotation about the axis 
${\bf A}_{m_1m}={\bf K}\times{\bf R}_{m_1m}$ and through the angle $2\vert{\bf
A}_{m_1m}\vert$. 
This is valid to quadratic order in ${\bf A}$. 
Explicitly,
\begin{equation}
\hat{D}_{m_1m}={\bf e}_z\times{\bf e}_R\circ{\bf e}_z\times{\bf e}_R
+\sin\vert 2{\bf A}\vert({\bf e}_R\circ{\bf e}_z-{\bf e}_z\circ{\bf e}_R)
+\cos\vert 2{\bf A}\vert({\bf e}_z\circ{\bf e}_z+{\bf e}_R\circ{\bf e}_R).
\end{equation}
The symbol ``$\circ$'' denotes the dyadic product and the orthogonal unit
vectors are ${\bf e}_z={\bf K}/K$ and
${\bf e}_R={\bf R}_{m_1m}/\vert{\bf R}_{m_1m}\vert$. Here, it is used that
${\bf K}$ is perpendicular to the plane, i.e.\ ${\bf K}\cdot{\bf R}=0$
for all vectors ${\bf R}$ within the plane.
Retaining in this expression only terms of up to second order in $K$ (since 
$A\ll 1$), one obtains
\begin{equation}
\hat{D}_{m_1m}=\hat{I}_3 +2\vert{\bf A}\vert
({\bf e}_R\circ{\bf e}_z-{\bf e}_z\circ{\bf e}_R)
-2\vert {\bf A}\vert^2({\bf e}_z\circ{\bf e}_z+{\bf e}_R\circ{\bf e}_R),
\end{equation}
where $\hat{I}_3$ is the identity matrix in three dimensions.
Note that ${\rm det}(\hat{D})=1+O(A^4)$, 
where $O(x)$ is Landau's big-O notation, i.e.\ that the length of the vector
$\boldsymbol{\rho}$ does not change (up to $O(A^4)$) due to the application
of the operator $\hat{D}$. Omitting the $A^2$-term, the length of the vector
would change in second order of $A$; then, $\hat{D}$ would not represent a
rotation.

The products of spin matrices in the three site diagrams are calculated
using the following scheme: After rotating the operators within the trace,
so that
$\hat{\rho}$ is the last operator, the spin factors preceding $\hat{\rho}$
in the trace are calculated in order to obtain an expression of the form
$f+g\boldsymbol{\sigma}$. Here, $f$ and $g$ are scalars in spin space. Thus,
taking the trace yields ${\rm Tr}\left((f+g\boldsymbol{\sigma})
\hat{\rho}\right)=f\rho+g\boldsymbol{\rho}$, an expression where the
spin operators have disappeared and are replaced by the particle density and
the spin orientation.
Using the operator identity
$e^{A}e^B\approx e^{A+B}e^{[A,B]/2}$, 
where the error is of third and higher order in the operators $A$ and $B$,
one obtains
\begin{equation}
e^{-{\rm i}\boldsymbol{\sigma}\cdot{\bf A}_{mm_2}}
e^{-{\rm i}\boldsymbol{\sigma}\cdot{\bf A}_{m_2m_1}}\approx
e^{-{\rm i}\boldsymbol{\sigma}\cdot{\bf A}_{mm_1}}
e^{{\rm i}\boldsymbol{\sigma}\cdot({\bf A}_{mm_1}\times{\bf A}_{mm_2})}
\label{aa}
\end{equation}
and 
\begin{equation}
e^{-{\rm i}\boldsymbol{\sigma}\cdot{\bf A}_{mm_2}}
e^{-{\rm i}\boldsymbol{\sigma}\cdot{\bf A}_{m_2m_1}}
e^{-{\rm i}\boldsymbol{\sigma}\cdot{\bf A}_{m_1m}}\approx
e^{{\rm i}\boldsymbol{\sigma}\cdot({\bf A}_{mm_1}\times{\bf A}_{mm_2})}.
\label{aaa}
\end{equation}
Using Eqs.\ (\ref{asigmaa}), (\ref{aa}), and (\ref{aaa}), which are all
valid to second order in $A$ (i.e.\ also in $K$), and Eq.\ (\ref{eisigma}), 
the spin structure of the three-site terms can be calculated, e.g.\ 
\begin{eqnarray}
{\rm Tr}\left(
e^{-{\rm i}\boldsymbol{\sigma}\cdot{\bf A}_{mm_2}}
e^{-{\rm i}\boldsymbol{\sigma}\cdot{\bf A}_{m_2m_1}}
e^{-{\rm i}\boldsymbol{\sigma}\cdot{\bf A}_{m_1m}}
\hat{\rho}\right)&\approx&
\cos\vert{\bf A}_2\vert\rho
+{\rm i}\frac{\sin\vert{\bf A}_2\vert}{\vert{\bf A}_2\vert}
{\bf A}_2\cdot\boldsymbol{\rho}\nonumber\\
&\approx&\rho+{\rm i}{\bf A}_2\cdot\boldsymbol{\rho},
\label{aproxaaa}
\end{eqnarray}
\begin{eqnarray}
{\rm Tr}\left(
e^{-{\rm i}\boldsymbol{\sigma}\cdot{\bf A}_{mm_2}}
\boldsymbol{\sigma}
e^{-{\rm i}\boldsymbol{\sigma}\cdot{\bf A}_{m_2m_1}}
e^{-{\rm i}\boldsymbol{\sigma}\cdot{\bf A}_{m_1m}}
\hat{\rho}\right)&\approx&
\hat{D}_{m_2m}\cdot\left[\cos\vert{\bf A}_2\vert\boldsymbol{\rho}
+{\rm i}\frac{\sin\vert{\bf A}_2\vert}{\vert{\bf A}_2\vert}{\bf A}_2\rho
-\frac{\sin\vert{\bf A}_2\vert}{\vert{\bf A}_2\vert}{\bf A}_2\times
\boldsymbol{\rho}\right]\nonumber\\
&\approx&\hat{D}_{m_2m}\cdot\boldsymbol{\rho}+{\rm i}{\bf A}_2\rho-{\bf
A}_2\times\boldsymbol{\rho},
\label{aproxasigmaa}
\end{eqnarray}
where the abbreviation ${\bf A}_2={\bf A}_{mm_1}\times{\bf A}_{mm_2}$ is
used, and the second approximation for each relation 
is the truncation to second order in
${\bf K}$.

The expressions (\ref{aproxaaa}) and (\ref{aproxasigmaa}) occur together
with the (complex) third order transition rate $W^{(3)}$. Adding the
corresponding complex conjugate diagram, one obtains some terms which contain
the real part $W^R=2{\rm Re}(W^{(3)})$ and some which contain the imaginary
part $W^I=2{\rm Im}(W^{(3)})$. Without SOI, only the real part is relevant,
since there is no complex pre-factor in this case.

\section{Third order transition probability for small polarons}
\label{appw3}

The third order diagrams (three-site hopping) give the additional
contribution to the
right-hand side of the rate equation (\ref{rate1})
\begin{eqnarray}
{\rm i}\sum_{m_1m_2}&\left\{{}-
e^{-{\rm i}\boldsymbol{\sigma}\cdot{\bf A}_{mm_2}}
e^{-{\rm i}\boldsymbol{\sigma}\cdot{\bf A}_{m_2m_1}}
\hat{\rho}_{m_1}
e^{-{\rm i}\boldsymbol{\sigma}\cdot{\bf A}_{m_1m}}
W^{(3)}_{m_1m_2m}
\right.&\nonumber\\
&{}+e^{-{\rm i}\boldsymbol{\sigma}\cdot{\bf A}_{mm_1}}
\hat{\rho}_{m_1}
e^{-{\rm i}\boldsymbol{\sigma}\cdot{\bf A}_{m_1m_2}}
e^{-{\rm i}\boldsymbol{\sigma}\cdot{\bf A}_{m_2m}}
W^{(3)*}_{m_1m_2m}
&\nonumber\\
&{}-\hat{\rho}_{m}
e^{-{\rm i}\boldsymbol{\sigma}\cdot{\bf A}_{mm_1}}
e^{-{\rm i}\boldsymbol{\sigma}\cdot{\bf A}_{m_1m_2}}
e^{-{\rm i}\boldsymbol{\sigma}\cdot{\bf A}_{m_2m}}
{W^{(3)*}_{mm_1m_2}}
&\nonumber\\
&\left.
{}+e^{-{\rm i}\boldsymbol{\sigma}\cdot{\bf A}_{mm_2}}
e^{-{\rm i}\boldsymbol{\sigma}\cdot{\bf A}_{m_2m_1}}
e^{-{\rm i}\boldsymbol{\sigma}\cdot{\bf A}_{m_1m}}
\hat{\rho}_{m}W^{(3)}_{mm_1m_2}
\right\}&.
\end{eqnarray}

The third order transition rates are $W^R_{m_1m_2m}=
W^{(3)}_{m_1m_2m}+W^{(3)*}_{m_1m_2m}=
2{\rm
Re}(W^{(3)}_{m_1m_2m})$ and $W^I_{m_1m_2m}=
-{\rm i}(W^{(3)}_{m_1m_2m}-W^{(3)*}_{m_1m_2m})=2{\rm
Im}(W^{(3)}_{m_1m_2m})$, derived from the quantity
\begin{equation}
W^{(3)}_{m_1m_2m}=\frac{1}{\hbar^3}J_{mm_2}J_{m_2m_1}J_{m_1m}I,
\end{equation}
where the symbol $I$ denotes the integral
\begin{eqnarray}
I({\bf E})&=&
e^{-3S_T}
\int_{-\infty}^{\infty}dt_1\int_0^{\infty}dt_2
\left(\exp\left\{
\sum_{\bf q}\frac{\Gamma_q}
{2\sinh(\hbar\omega_q\beta/2)}\right.\right.\nonumber\\
&\times&\left.\left.
\left[
\cos\omega_q(t_1+{\rm i}\hbar\frac{\beta}{2})
+\cos\omega_q(t_2+{\rm i}\hbar\frac{\beta}{2})
+\cos\omega_q(t_1-t_2+{\rm i}\hbar\frac{\beta}{2})
\right]
\right\}-1\right)\nonumber\\
&\times&\exp\left\{
\frac{\rm i}{\hbar}e{\bf E}\cdot{\bf R}_{mm_1}t_1
-\frac{\rm i}{\hbar}e{\bf E}\cdot{\bf R}_{mm_2}t_2
\right\}
\label{w3int}
\end{eqnarray}
with the abbreviations
\begin{equation}
\Gamma_q=\frac{\vert\gamma(q)\vert^2(1-\cos{\bf q}\cdot{\bf g})}{N}
\end{equation}
and
\begin{equation}
S_T=\sum_{\bf q}\frac{\Gamma_q}{2}\coth(\hbar\omega_q\frac{\beta}{2}).
\end{equation}
The sums runs over the phonon wave-vectors ${\bf q}$. $\gamma(q)$ is the
electron-phonon interaction constant, $\omega_q$ the phonon frequency, $N$
the number of phonon states,
$\beta=1/(k_BT)$, and ${\bf g}$ is a vector connecting two nearest-neighbor
sites. Note, that the integral is independent of the direction of ${\bf g}$.
Thus, the resulting expression is isotropic.

Taking two times the real part, the integral over $t_2$ would extend from
$-\infty$ to $\infty$. This would be the expression to be calculated for the
small polaron Hall effect.\cite{FriedmanHolstein63,BoettgerBryksin85} Here,
we cannot take this step, because we also need to calculate the imaginary
part of the expression (\ref{w3int}).

The three-site rate obeys the principle of detailed balance
\begin{equation}
W^{(3)}_{mm_2m_1}=W^{(3)}_{m_1m_2m}e^{-\beta e{\bf E}\cdot{\bf R}_{mm_1}}
\end{equation}
and has the further symmetry
\begin{equation}
W^{(3)}_{m_1mm_2}=W^{(3)}_{m_1m_2m}.
\end{equation}
Using these symmetries, the long wave-length limit of the third order term
Eq.\ (\ref{vecwriq}) for a triangular lattice (hexagonal crystal symmetry)
can be calculated as
\begin{equation}
{\bf W}^{R}({\bf q})={\rm i}{\bf K}\circ{\bf q}\cdot({\bf E}\times{\bf K})
\frac{3a^4e}{k_BT}{\rm Re}(W_{012}^{(3)}).
\label{wrofq}
\end{equation}
In order to calculate ${\bf W}^I$ the imaginary part of $W^{(3)}$ has to be
used instead of the real part. The indices $0$, $1$ and $2$ denote the sites
located at the vertices of an elementary triangle of the lattice.
$W^{(3)}_{012}$ has to be calculated in zero electric field. Thus, there only
remains to determine the integral $I({\bf 0})$ in order to obtain the third
order coefficients.

For a sufficiently large phonon dispersion, the main contribution to the
integral $I$ comes from a saddle point $(t_{1s},t_{2s})$ at 
$({\rm i}2\hbar\beta/3,{\rm i}\hbar\beta/2)$.
Under the condition $\hbar\omega_q\ll k_BT$ (high temperatures compared
with the Debye temperature), the exponent in the integrand 
can be expanded in a series in $t_1$ and $t_2$, which is truncated after the
second order terms. The integral can then be calculated and one obtains
\begin{equation}
I\approx\frac{\hbar^2\pi}{E_ak_BT4\sqrt{3}}e^{-\frac{4E_a}{3k_BT}}
\left(1-{\rm i}\sqrt{\frac{3k_BT}{E_a\pi}}e^{\frac{E_a}{3k_BT}}
\right),
\label{itlarge}
\end{equation}
where
\begin{equation}
E_a=\sum_{\bf q}\frac{\hbar\omega_q\Gamma_q}{4}.
\end{equation}

Combining Eq.\ (\ref{itlarge}) with Eq.\ (\ref{wrofq}), the mobility
$\mu_{yx}$ connected with
the anomalous Hall effect and the spin Hall effect reads
\begin{equation}
W^RK^2=\mu_{yx}=\mu_0\frac{\sqrt{3}\pi}{4}\frac{(Ka)^2J^3}{E_a(k_BT)^2}
e^{-\frac{4E_a}{3k_BT}},
\end{equation}
where $\mu_0=ea^2/\hbar$ and $a$ is the distance between nearest 
neighbors (lattice constant).
This expression can be compared with the
Hall mobility $\mu_H$ 
for small polaron hopping\cite{BoettgerBryksin85} to give the
relation
\begin{equation}
\frac{\mu_{yx}}{\mu}=2(Ka)^2\frac{\mu_H}{\mu_0}
=\frac{2\hbar K^2\mu_H}{e}.
\label{muyxvsmuh}
\end{equation}

The effective activation energy for the $W^R$ processes (e.g.\ the spin
Hall effect) is $4E_a/3$ [see the real part of Eq.\ (\ref{itlarge})], 
which is identical to the situation for
the ordinary Hall effect. On the other hand, the activation energy
for the $W^I$ processes is $E_a$, i.e.\ $W^I\gg W^R$ for small temperatures
$k_BT\ll E_a$.

\section{Magnetic fields due to spin accumulation and electric current}
\label{appmag}

The magnetic field generated by the spin polarization (magnetization) 
${\bf M}({\bf r})={\bf M}(r_y)=\mu_B\boldsymbol{\rho}(y)$ in the strip reads
\begin{equation}
{\bf H}_M(y,z)=-K\boldsymbol{\nabla}\frac{1}{2\pi}\int_{-b}^{b}dy^\prime
\frac{M_y(y^\prime)
(y-y^\prime)+M_z(y^\prime)z}{z^2+(y-y^\prime)^2}.
\end{equation}
Note, that the field ${\bf H}_M$ has no $x$-component, even though the spin
polarization has a finite $x$-component $M_x$.
The magnetic field ${\bf H}_M$ may be exploited in order to detect the spin
accumulation. In this case, one has to keep in mind, that the electrical
current in $x$-direction also leads to a magnetic field 
${\bf H}_I$.\cite{Hirsch99} Both
fields have different symmetry. Thus, if ${\bf H}_M$ is not very much
smaller than ${\bf H}_I$, a measurement of the magnetic field near the
strip allows a detection of the spin accumulation.

Explicitly, the magnetic field generated by the (homogeneous) current 
$I=eb\mu E\rho$ within the strip reads
\begin{equation}
{\bf H}_I(y,z)=\frac{I{\bf e}_y}{4\pi b}\arctan\left(\frac{2az}{y^2+z^2-b^2}
\right)+\frac{I{\bf e}_z}{8\pi b}\log\left(\frac{z^2+(y-b)^2}
{z^2+(y+b)^2}\right).
\end{equation}


\begin{figure}[t]
\includegraphics*{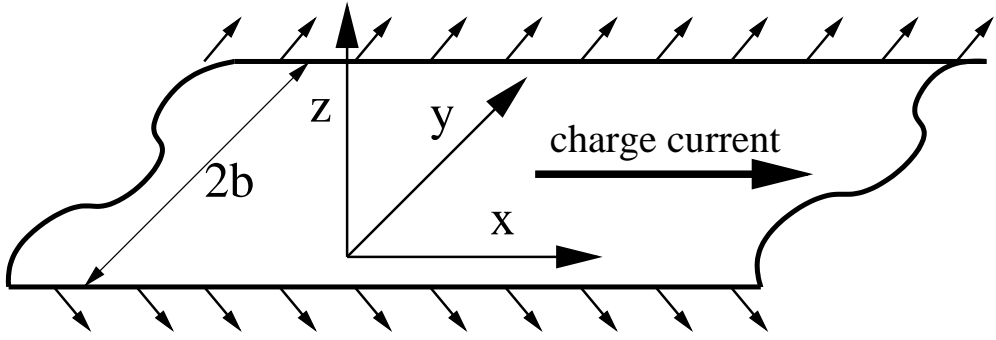}
\caption{The co-ordinate system chosen for the strip of finite width. 
The charge
current within the strip is non-polarized. But at the transversal boundaries 
$y=\pm b$, a
finite spin density occurs due to $\epsilon_R$, i.e.\ this situation leads
to spin accumulation.}
\label{figstrip}
\end{figure}

\begin{figure}[t]
\includegraphics*[angle=-90,width=0.8\textwidth]{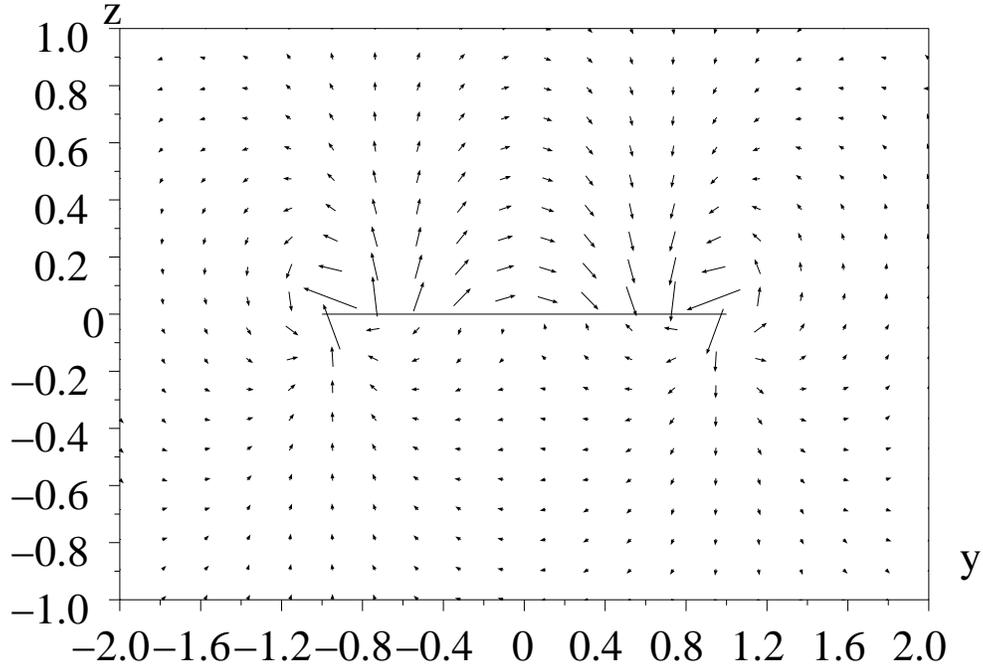}
\caption{The magnetic field due to the spin accumulation in the
$z$-$y$-plane. The parameters are 
$b=1$, $\epsilon=1$. The horizontal line denotes the extent of the strip.}
\label{fighm}
\end{figure}

\begin{figure}[t]
\includegraphics*[angle=-90,width=0.8\textwidth]{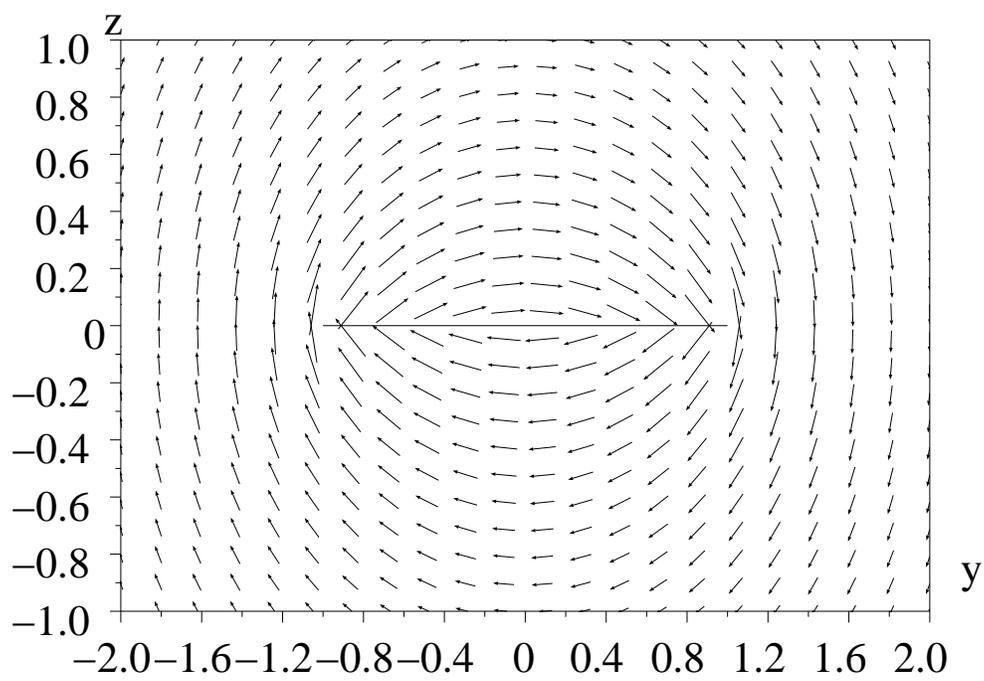}
\caption{The magnetic field due to the current through the strip. The
parameters are the same as in Fig.\ \ref{fighm}.}
\label{figcurrent}
\end{figure}

\end{document}